\documentclass[twocolumn,amsmath,amssymb,%
               superscriptaddress]{revtex4}
\usepackage{graphicx}
\usepackage{dcolumn}
\usepackage{bm}

\begin{document}

\newcommand{\beq}{\begin{equation}}
\newcommand{\eeq}{\end{equation}}

\title{Stability of Kramers Majorana doublets: the effects of interactions and disorders}
\author{Xiao Xiao}
\affiliation{Department of Physics, The Hong Kong University of Science and Technology, Hong Kong, the People's Republic of China}
\date{\today}
\begin{abstract}
  In this work we study the effects of interactions and disorder on $1D$ DIII topological superconductors and the Majorana Kramers doublets (MKDs). In contract to the case without the time-reversal symmetry, the Umklapp interaction plays important roles in this system. The underlying phases due to the Umklapp interaction and disorder are found by using a perturbative renormaliztion analysis based on the Abelian Bosonization. Importantly, the stable topological regime can be found within a rather wide parameter space. Furthermore, the degeneracy splitting of the MKDs is shown to be still exponentially dependent on the length of the wire in the presence of both the Umklapp interaction and disorder, when the Luttinger parameter $K_0>\sqrt{2}/2$. The differences caused by the Umklapp interaction are highlighted in contrast to the time-reversal breaking cases.
\end{abstract}
\maketitle

\begin{center}
\textbf{Introduction}
\end{center}

Owning to its exotic non-Abelian braiding statistics \cite{Ivanov01} and its potential applications in quantum computations \cite{Kitaev03,Nayak08,Alicea11}, Majorana zero-modes have been intensively pursued since their existence had been demonstrated in the $p+ip$ superconductors in $2D$ \cite{Read00,Gurarie05,Stone06} and also in the $p$ wave superconductors in $1D$ \cite{Kitaev01}. Recently many theoretical proposals, such as topological insulator/superconductor structures \cite{Fu08,Nilsson08,Fu09,Knez12}, semiconductor-superconductor heterostructures \cite{Sau10,Lutchyn10,Oreg10,Cook11,Mourik12,Rokhinson12,Das12,Deng12a,Finck12,Albrecht16,Chen17}, and magnetically-ordered metallic systems coupled to an s-wave superconductor \cite{Choy11,Perge13,Klinovaja13,Braunecker13,Vazifeh13,Pientka13,Perge14}, have been put forward to realize topological superconductors, from the boundary or the defects of which the Majoran zero-modes are emergent. In most of proposals for $1D$ systems, the time-reversal symmetry is broken, and the realized topological superconductors (TSCs) are belong to the D class \cite{Altland97,Schnyder08,Kitaev09}, in which TSCs usually supports an unpaired Majorana mode at the ends.

However, when the time-reversal symmetry is restored, a $1D$ TSC belong to the DIII class \cite{Schnyder08,Kitaev09} can be realized. In this kind of TSC, a pair of Majorana zero modes protected by time-reversal symmetry can emerge at its boundaries, so the pair of Majorana modes is denoted as the Majorana Kramers doublet (MKD). The same with its unpaired counterpart, the MKD also shows very interesting non-Abelian braiding statistics and would be thus useful for the topological quantum computations either \cite{Liu14,Gao16}. This kind of TSC preserving the time-reversal symmetry may be realized by a Josephson $\pi$-junction mediated by the helical edge modes of a quantum spin Hall insulator \cite{Fu09,Keselman13,Schrade15} or the proximity of a Rashba nanowire to unconventional superconductors \cite{Wong12,Deng12b,Zhang13,Nakosai13}. Recently, transport signatures of MKDs and its detection method was proposed by using a quantum point contact in a quantum spin Hall system \cite{Li16}. Moreover, other recent works investigated the transport signatures of MKDs in junctions \cite{Kim16}, the Kondo effect of MKDs \cite{Bao16} and Josephson effects of MKDs \cite{Schrade18}. These may be important steps toward the observation of MKDs in experiments.

On the other hand, in real experiments on semiconducting nanowires the effect of disorder is usually hard to be avoid. A lot of previous studies suggested that the disorder has profound influence on the $1D$ topological superconductors in the D class \cite{Potter10,Brouwer11a,Brouwer11b,Lutchyn11,Stanescu11,Sau11,Liu12}. For example, the disorder would change the ground state degeneracy splitting of Majorana modes in a $1D$ D class topological superconductor from an exponential to an algebraic dependence on the length of wire \cite{Brouwer11a}. As a consequence, increasing the disorder should drive a quantum phase transition from a TSC phase supporting Majorana end states to a trivial phase without Majorana modes. More importantly, the low-energy properties of quasi-$1D$ systems can be dramatically affected by the interplay of the disorder and interaction \cite{Giamarchi88,Lobos12}. Therefore, it would be very experimentally relevant to study how the combination of disorder and interaction affects the stability of a $1D$ DIII TSC and the MKDs harbored therein.

\begin{figure}[!th]
  \centering
  \includegraphics[width=1\columnwidth]{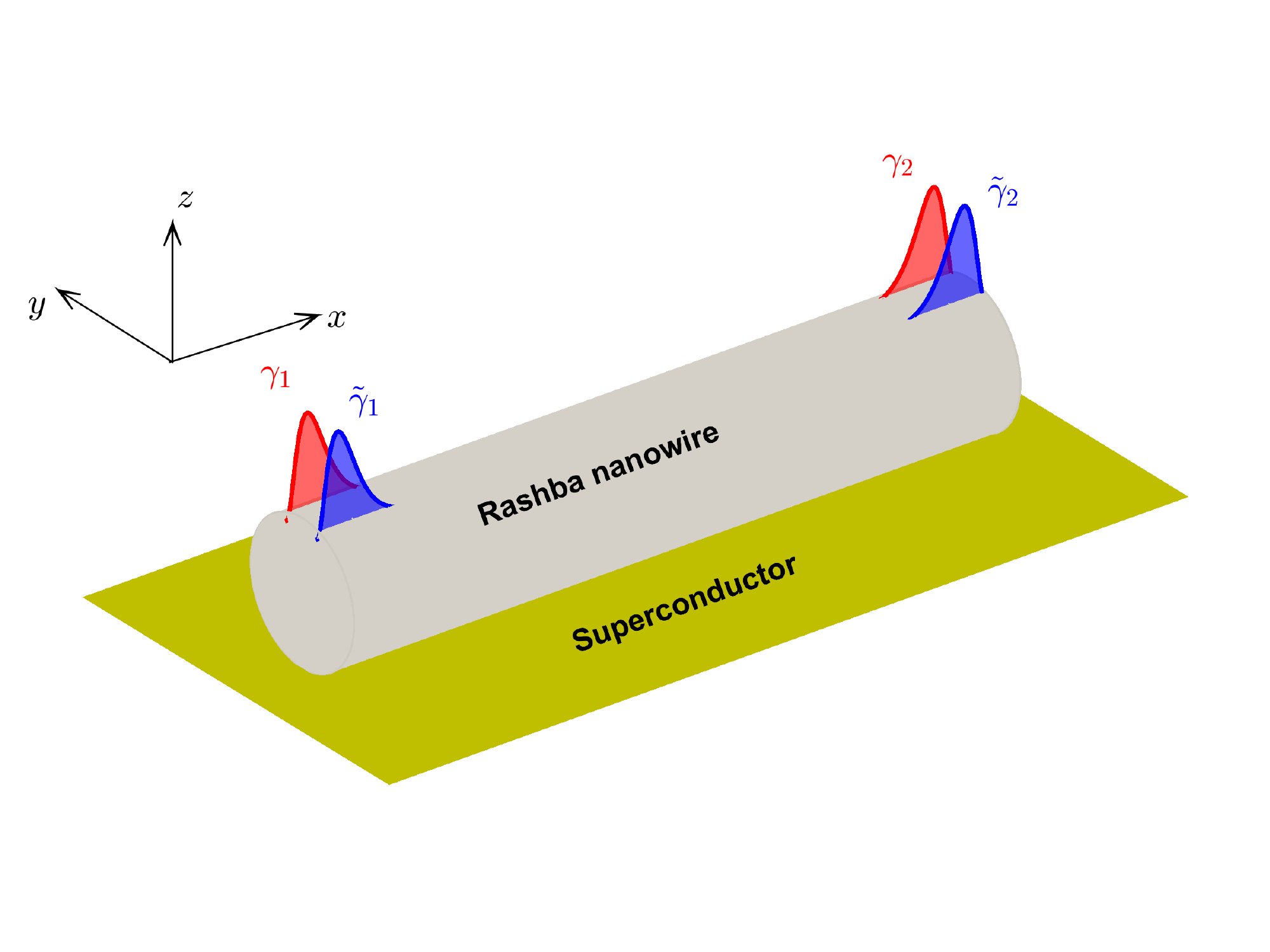}
  \caption{The schematic illustration of a $1D$ topological superconductor in the DIII class. The Majorana Kramers doublet protected by the time-reversal symmetry locates at the ends of the topological superconductor.}
   \label{fig:urtraj1}
\end{figure}

In this work, we investigate the effects of interactions and disorder on a $1D$ DIII TSC by constructing a low-energy effective theory based on the Bosonization technique. Interestingly, when the interactions and disorder are absent, the Bosonization analysis indicates that the superconducting gaps from the two pairs of Rashba bands should have opposite signs to make the MKD survive, or the two Majorana fermions at the ends of the wire would annihilate each other to form a Fermion. Therefore, the DIII TSCs would be realized, when the chemical potential locates at the band center such that the Fermi wave vector fulfills $k_Fa \in [\pi-q_0a,\pi+q_0a]$ with $q_0a$ determined by the ratio between Rashba spin orbit coupling and the hopping. The position of chemical potential thus implies that the Umklapp interaction would play important roles \cite{Gogolin98,Giamarchi04}. When the interactions and disorder are turned on, a group of coupled renormalization group (RG) equations is obtained based on the low-energy effective theory. By the analysis of the RG flows, the underlying phase diagram is determined. Interestingly, the topological superconducting phase is found to be stable even in the presence of repulsive interactions (with Luttinger parameter $K<1$) with finite Umklapp interaction and disorder. Further the degeneracy splitting of MKD is analyzed by using an instanton argument. The exponential dependence of the degeneracy splitting on the wire length can persist, as long as the strengths of the Umklapp interaction and disorder are small in comparison with that of the superconducting order. Importantly, the degeneracy splitting could have much higher tolerance for the Umklapp interaction and disorder, when the Luttinger parameter $K_0>\sqrt{2}/2$. Given to the recent success of observing perfect quantization due to the Majorana in the Rashba nanowire system with time-reversal breaking \cite{Zhang17}, our study indicates the possibility of realizing a $1D$ DIII class TSC and the MKDs harbored therein in the wire geometry.

\begin{center}
\textbf{Low-energy effective theory}
\end{center}

Generically the Kramers Majorana doublet can be achieved by the proximity of a Rashba nanowire to a unconventional superconductor. The minimal Hamiltonian may be written as the following tight-binding form:
\begin{align}
H = &-t \sum_{\langle i,j \rangle,\alpha} c_{i,\alpha}^{\dag} c_{j,\alpha} - i\lambda_R \sum_{\langle i,j \rangle} c_{i,\alpha}^{\dag} \left(\vec{\sigma}^{\alpha\beta} \times \vec{d}_{ij}\right)_z c_{j,\beta} \nonumber \\
&+\Delta \sum_{\langle i,j \rangle} \left(c_{i,\uparrow}^{\dag}c_{j,\downarrow}^{\dag} + h.c.\right) - \mu\sum_{j,\alpha} c_{j,\alpha}^\dag c_{j,\alpha},
\end{align}
where $t$ is the hopping between the nearest neighbor sites labeled by $\langle i,j \rangle$ in the summation, $\lambda_R$ is the strength of Rashba spin-orbit coupling (SOC), $\Delta$ is the strength of superconducting pairing, and $\mu$ is the chemical potential. It can be shown that the Hamiltonian density $\mathcal{H}$ fulfills both the time-reversal and particle-hole symmetries with the two operators defined as $\mathcal{T}=\tau_0\otimes i\sigma_y \mathcal{K}$  and $\mathcal{P} = \tau_x\otimes \sigma_0 \mathcal{K}$ with $\mathcal{K}$ denoting the complex conjugate operator. In the above, $\tau$-matrices and $\sigma$-matrices act on the particle-hole and the spin space respectively. Then such a system belongs to the DIII class according to the classification of non-interacting topological phases.

Around the Fermi level, the low-energy effective Hamiltonian becomes:
\begin{align}
H = &-iv_F \sum_{\eta=\pm} \int dx \left[R_{\eta}^\dag(x)\partial_xR_{\eta}(x)-L_{\eta}^\dag(x)\partial_xL_{\eta}(x)\right] \nonumber \\
& + 2\Delta \cos(k_F+q_0)a \int dx \left[L_+^\dag(x) R_-^\dag(x) + h.c.\right] \nonumber \\
& + 2\Delta \cos(k_F-q_0)a \int dx \left[R_+^\dag(x) L_-^\dag(x) + h.c.\right],
\end{align}
where $O_{\eta}^\dag$ and $O_\eta$ are the creation and annihilation operators for the left-moving ($O=L$) and right-moving $O=R$ quasi-particles in the $\eta$-band, $v_F=2\tilde{t}a\sin(k_Fa)$ with $\tilde{t}=\sqrt{t^2+\lambda_R^2}$, $q_0a=\arctan(\lambda_R/t)$, and $k_F$ is the Fermi wave vector.

In the experimental relevant situations, both \textit{e-e} interactions and disorders may play important roles. Since now the regime of interesting is spin-full, the \textit{e-e} interaction can be classified by the so-called ``g-ology'', namely the backward scattering $g_1$, dispersive scattering $g_2$, Umklapp scattering $g_3$ and forward scattering $g_4$. The back scattering $g_1$ is known to be irrelevant even in the presence of SOC, and we may ignore it in the latter discussion \cite{Schulz09}. For both the dispersive $g_2$ and forward scattering $g_4$, the momentum transfered in the scattering is $q\sim0$, so we would have $g_2=g_4=V(q\sim0)=g$ with $V(q)$ the \textit{e-e} interaction in momentum space. It would be clear from the following discussion that the Umklapp scattering $g_3=g_u$ with the momentum transfer $q\sim\pi$ should be important for our interest, and we will take this term into account. On the other hand, the impurities in the nanowires may be described by the quenched disorder with a short-range Gaussian disorder potential $V(x)$ characterized by the correlation $\langle V(x) V(y) \rangle=D\delta(x-y)$. The disorder Hamiltonian is given by $H_{im} = \int dx V(x) \rho(x)$ with $\rho(x)$ the fermionic density at position $x$ of the nanowire.

Then gathering these factors and adopting the usual Bosonization transformation $O_{\eta}(x) = \frac{\mathbb{K}_{O,\eta}}{\sqrt{2\pi\alpha}}e^{iu_{O}2\phi_{O,\eta}(x)}$, the bosonized low-energy effective Hamiltonian can be obtained. In the above $\mathbb{K}_{O,\eta}$ denotes the Klein factors, and $\phi_{O,\eta}(x)$ represents a Bosonic field with $O=L,R$ denoting left-moving and right-moving components. We note that it would be convenient to work in a new set of bosonic fields, which relate with the original ones $\{ \phi_{O,\eta}(x) \}$ as:
\beq
\begin{cases}
\phi_{R,+}=\frac{1}{\sqrt{2}}(\varphi_1+\vartheta_1),~\phi_{L,+} = \frac{1}{\sqrt{2}}(\varphi_2+\vartheta_2), \\
\phi_{R,-}=\frac{1}{\sqrt{2}}(\varphi_2-\vartheta_2),~\phi_{L,+} = \frac{1}{\sqrt{2}}(\varphi_1-\vartheta_1).
\end{cases}
\eeq
The low-energy effective action in terms of bosonic fields $\{\varphi_{j},\vartheta_{j}\}$ can be written as following by treating the quenched disorder with the usual replica technique (see App.A for details):
\begin{widetext}
\begin{align}
S &= \sum_{\mu}\int dx d\tau \left[-\left(\mathcal{L}_1^\mu + \mathcal{L}_2^\mu\right) + \frac{\Delta\cos(k_F+q_0)a}{\pi \alpha} \sin2\vartheta_1^\mu + \frac{\Delta\cos(k_F-q_0)a}{\pi \alpha} \sin2\vartheta_2^\mu + \frac{g_u}{2\pi^2 \alpha^2} \cos 2(\varphi_1^\mu(x) + \varphi_2^\mu(x))\right] \nonumber \\
&-\frac{D}{\pi^2 \alpha^2} \sum_{\mu,\nu} \int dx d\tau d\tau' \left[\begin{array}{c} \cos\left(\vartheta_1^\mu(x,\tau)+\vartheta_2^\mu(x,\tau)\right) \cos\left(\vartheta_1^\nu(x,\tau')+\vartheta_2^\nu(x,\tau')\right) \\ \cos\left(\varphi_1^\mu(x,t)+\varphi_2^\mu(x,t)-\varphi_1^\nu(x,\tau')-\varphi_2^\nu(x,\tau')\right) \end{array}\right],
\end{align}
\end{widetext}
where $\mu$ and $\nu$ are the replica indices, $\mathcal{L}_{j}$ is the Lagrangian density for a spinless Luttinger liquid:
\beq
\mathcal{L}_{j} = \frac{i\partial_x\vartheta_i}{\pi} \partial_\tau \varphi_i - \frac{u}{2\pi K}(\partial_x \varphi_i)^2 - \frac{uK}{2\pi} (\partial_x \vartheta_i)^2,
\eeq
where the Luttinger parameter $K = \frac{1}{\sqrt{1+g/\pi v_F}}$ depends on the properties of interaction $g$, and $u=v_F/K$. It is obvious that the bosonic fields fulfills the following commutation relation: $[\varphi_j(x),\vartheta_k(y)]=i\pi/2 \textrm{Sgn}(y-x) \delta_{j,k}$.

When the effects of the Umklapp interaction and disorders are absent, the effective action reduces to two copies of Sine-Gordon action with each copy describing a Kitaev chain \cite{Lobos12,Fidkowski11}. When $K=1$, $\Delta$ would flow to the strong coupling. Under this situation, the terms proportional to $\sin\vartheta_{i}$ should be pinned to its minimum. Supposing that $\cos(k_F+q_0)a$ and $\cos(k_F-q_0)a$ have the same sign, the two fields $\vartheta_1$ and $\vartheta_2$ are thus pinned to the same value up to a global $Z_2$ transformation $\vartheta_i\rightarrow\vartheta_i+\pi$. In this case, the two Majorana at one end of the wire would be self-conjugate with each other and annihilate to a Fermion. Therefore, the existence of Majorana Kramers doublets requires that $k_Fa$ should be within $[\pi-q_0a,\pi+q_0a]$ (near half-filling), so that $\vartheta_1$ is pinned to either of the two degenerate minima $\vartheta_1=\{\pi/4,-3\pi/4\}$, while $\vartheta_2$ is pinned to either of $\vartheta_2=\{-\pi/4,3\pi/4\}$. Then the minima of $\vartheta_1$ relates to those of $\vartheta_2$ as $\vartheta_1=-\vartheta_2$ up to a global $Z_2$ transformation, which is a manifest of time-reversal symmetry. As far as we are interested in the regime where a topological superconducting phase supports Majorana Kramers doublets, the system is just around the half-filling. Therefore, it is necessary to take into account the Umklapp interaction (the $g_u$ term in Eq.(4)). Since $q_0 a$ is typically much smaller than $\pi$ in experiments, we will set $\tilde{\Delta}=\pm \Delta\cos(k_F\pm q_0)a$ in the following discussion.

\begin{figure}[!th]
  \centering
  \includegraphics[width=1\columnwidth]{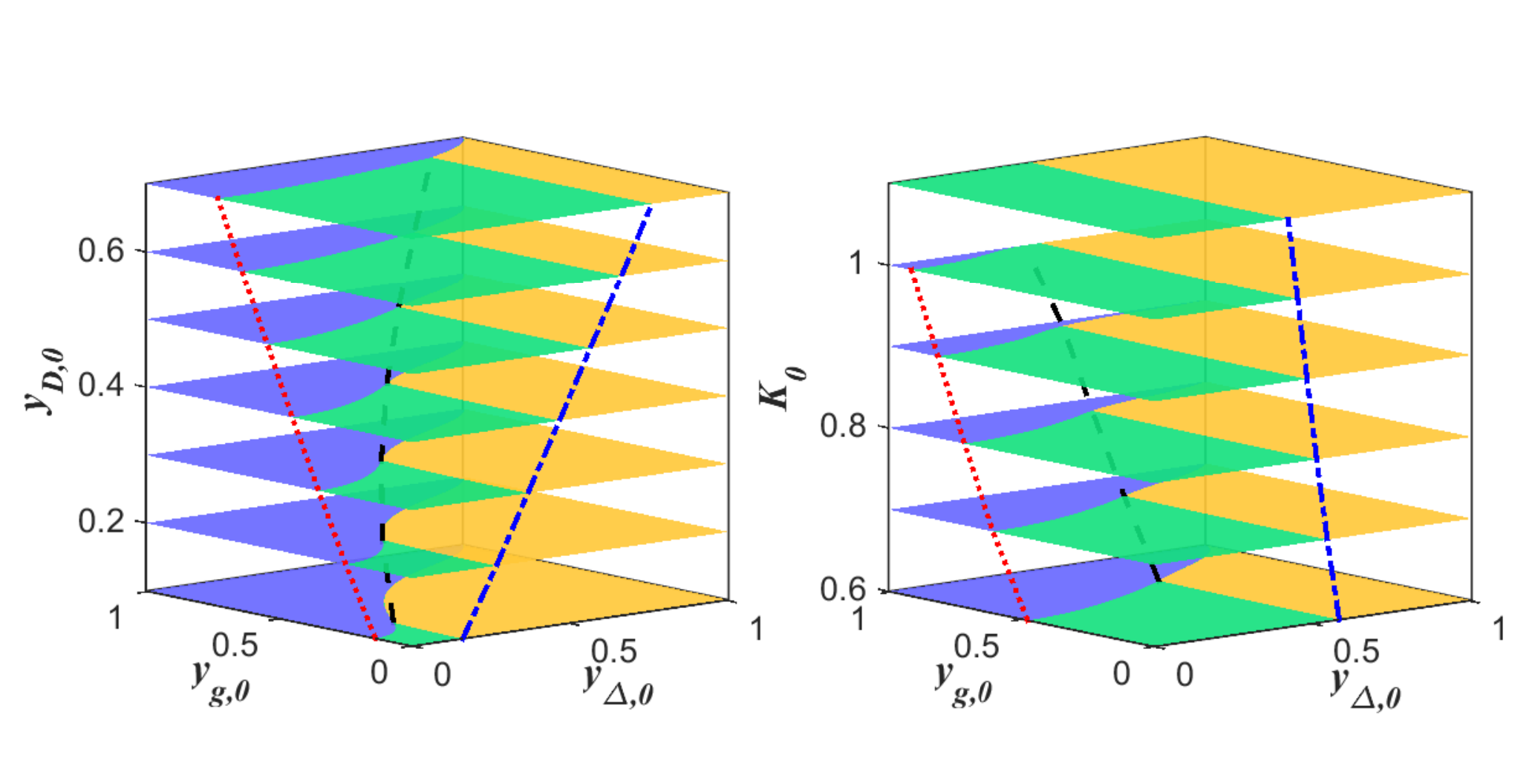}
  \caption{The phase diagrams determined by the analysis of the RG equations: (a) with fixed Luttinger parameter $K=0.65$; (b) with fixed disorder strength $y_{D}=0.5$. In the figures, the purple region is for the CDW-I phase with $y_g$ dominating, the green region is for the CDW-II phase with $y_D$ dominating, and the yellow region is for the TSC phase with $y_\Delta$ denominating. The curves are to denotes the evolution of the phase boundaries.}
   \label{fig:urtraj2}
\end{figure}

\begin{center}
\textbf{The effects of interaction and disorder}
\end{center}

To understand the possible phases and how they are controlled by interactions and disorders, we applied the standard perturbative renormalization group (RG) analysis. We notice that the action is invariant, when we exchange the fields labelled by $1$ with those labelled by $2$. Thus the renormalization can be performed by just considering the correlation function:
\beq
R_{\vartheta_1}(x_2-x_1,\tau_2-\tau_1) = \left\langle T_{\tau} e^{i\vartheta_{1}(x_1,\tau_1)} e^{-i\vartheta_{1}(x_2,\tau_2)} \right\rangle.
\eeq
By using the standard procedures, we found the following RG flow equations (see App.B for details):
\begin{align}
\frac{dy_\Delta}{d\ell} &= \left(2-K^{-1}\right)y_\Delta,  \\
\frac{dy_g}{d\ell} &= \left(2-2K\right)y_g,  \\
\frac{dy_D}{d\ell} &= \left(3-(K^{-1}+K)\right)y_D,  \\
\frac{dK}{d\ell} &= y_\Delta^2 - \frac{y_g^2 K^2}{4} - \frac{y_D}{8}\left(K^2-1\right),   \\
\frac{du}{d\ell} &= \frac{y_D}{8}\left(K-\frac{1}{K}\right) u,
\end{align}
where the dimensionless parameters are defined as: $y_\Delta = \tilde{\Delta}\alpha/v$, $y_g = g_u/\pi u$, and $y_D = D\alpha/\pi u^2$.

We notice that Eq.(7) is just the RG flow equation for a Kitaev chain \cite{Lobos12,Fidkowski11}. When the interaction is attractive or even weak repulsive ($K>1/2$), the superconducting order increases with the length of the wire and flow to the strong coupling. For the case of Kitaev chain, the field $\vartheta$ is pinned to one of the minima, and the global $Z_2$ symmetry is broken spontaneously. For the MKD case here, $\vartheta_1$ and $\vartheta_2$ are pinned to the minima connected by a time-reversal transformation with $\vartheta_1=-\vartheta_2$. The Umklapp interaction couples with the dual fields $\varphi_j$, so it would complete with the superconducting order. Indeed Eq.(8) indicates that the Umklapp interaction favors the repulsive interaction with $K<1$. Since how $y_\Delta$, $y_g$ and $y_D$ is renormalized is determined by the value of Luttinger parameter $K$ (see Eq.(7)-(9)), a more transparent way to see the competition between these terms is to check how they renormalize the Luttinger parameter $K$ (see Eq.(10)). Along the renormalization procedures, $y_{\Delta}$ drives the Luttinger parameter $K$ to flow to a larger value (more attractive), while $y_g$ makes $K$ to flow to a smaller value (more repulsive). From Eq.(10), we also notice the interesting role of disorder, which always tries to help the `weaker' one. For example, when $K>1$, $y_{\Delta}$ tries to flow to the strong coupling, but $y_D$ tries to reduce the value of $K$ through renormalization. In this way, $y_D$ competes with $y_\Delta$. On the other hand, when $K<1$, $y_D$ competes with $y_g$. The above analysis suggests that the non-interacting limit $K=1$ is not stable, and the disorder and interaction can drive the system to different phases.

To get more transparent picture about the competition between these factors, we numerically analyzed the RG flow equations. Due to the perturbation nature of the approach, the integration of the RG flow equations should stop at the length scale $\ell^*$, where one of the coupling strengths arrive at the strong coupling $\max\{y_\Delta,y_g,y_D\}=1$. It is obvious that there is no fixed point for the flow equations, so the underline phase should be determined by the coupling arriving at the strong coupling limit first \cite{Lobos12}. The phase dominated by $y_{\Delta}$ is the topological superconducting phase supporting Majorana Kramers doublets (denoted by TSC), the phase with $y_{g}$ flowing to the strong coupling first would pin the field $\varphi_1+\varphi_2$ and corresponds to a charge density wave phase formed by both the $+$ and $-$ bands (denoted as CDW-I), while the phase with $y_{D}$ arriving at the strong coupling first also corresponds to a charge density order but formed from either the $+$ or $-$ band (denoted as CDW-II). To determine the various phases driven by the interaction and disorder, we begin from a set of initial coupling strengths $\{y_{\Delta,0},y_{g,0},y_{D,0},K_0\}$ and track how these coupling strengths flow with the RG process. Then the phase diagrams in the parameter space spanned by $\{y_{\Delta,0},y_{g,0},y_{D,0},K_0\}$ can be obtained. In Fig.~2(a), we fixed the initial Luttinger parameter as $K_0=0.65$ and determined the phases in the parameter space spanned by $\{y_{\Delta,0},y_{g,0},y_{D,0}\}$. As one expected, with the increment of $y_{D,0}$, the regime of CDW-II increases. With this initial $K_0$, all the three phases can be reached by tuning the parameters $\{y_{\Delta,0},y_{g,0},y_{D,0}\}$. In Fig.~2(b), we fixed $y_{D,0}=0.5$ and determined the phases in the parameter space $\{y_{\Delta,0},y_{g,0},K_{0}\}$. We found that the CDW-I phase would disappear, when the \textit{e-e} interaction is attractive with $K_0>1$, which is consistent with the fact that $y_g$ favors the repulsive interactions. Importantly, the TSC regime can exist even when the interaction is repulsive, and the regime is enlarged as the interaction become more attractive.

\begin{center}
\textbf{The stability of MKDs}
\end{center}

The above analysis indicates that the TSC phase supporting MKD can exist even in the presence of repulsive interaction and disorder. In this part, we further analyze whether the MKD will be destroyed by interaction and (or) disorder. To study the stability of MKD against interaction and disorder, we need to calculate the energy splitting of the zero-energy modes $\delta E \propto \exp(-S_{inst})$ in the TSC phase, where $y_{\Delta}$ flows to the strong coupling first. In the absence of the Umklapp interaction and disorders, the instanton action is given by:
\beq
\mathcal{S}_{inst} = \frac{8\sqrt{K}}{\pi} \frac{L}{\xi}.
\eeq
As it is expected from Eq.(4), the instanton action is just twice of that for a Kitaev's chain \cite{Lobos12,Fidkowski11}.

The interaction and disorder give contributions to $\mathcal{S}_{inst}$ in two ways: \textit{i}) the contribution explicitly from the terms in the action specified in Eq.(4); \textit{ii}) the indirect influence of $y_g$ and $y_D$ on the Luttinger parameter $K$ through the RG flow. To calculate these contributions, the RG equations have to be integrated to $\ell^*=\ln(y_{\Delta,0})/(K_0^{-1}-2)$ within the lowest order approximation $K(\ell)\approx K_0$, and then the instanton action $\mathcal{S}_{inst}$ can be calculated in the presence of interaction and disorder at $\ell=\ell^*$. We first consider the explicit contribution \textit{i}). Within the TSC regime, we can evaluate the contributions from $y_g$ and $y_D$ perturbatively. For the Umklapp interactions (the term $\sim y_g$), due to the strongly fluctuating properties of field $\varphi_1$ and $\varphi_2$, the contribution to $\mathcal{S}_{inst}$ vanishes in the first order of $y_g$. On the other hand, the contribution from the disorder effect $\sim y_D$ is nonzero and can be evaluated to be $\sim\exp(-\xi/\ell_e)$, where $\ell_e$ is the scattering length (see Appendix C for details). This term is sub-leading in the thermodynamic limit $\xi/L\rightarrow0$. As it is expected, the results above is very similar with those of a Kitaev's chain \cite{Lobos12}, because without the Umklapp contribution the problem reduces to the effect of disorder on two copies of Kitaev chains.

Then we consider the indirect contribution \textit{ii}) through the RG flow. In this case we just need to replace the Luttinger parameter $K$ in Eq.(12) by $K(\ell^*)$. It would be convenient to introduce the scattering length $l_e$ and the length scale associated with Umklapp interaction $l_u$ (see App.C for the detailed definition). At the length scale $\ell^*$, we found that the Luttinger parameter $K(\ell^*) = K_r - \delta K_g - \delta K_D$, where $K_r = K_0 + \frac{K_0}{4K_0-2}$ is the renormalized Luttinger parameter in the absence of Umklapp interaction and disorders, $\delta K_g = \mathcal{C}(K_0) (k_Fl_u)^{-2} (k_F\xi)^{2\nu_1}$ is the correction due to the Umklapp interaction, and $\delta K_D = \mathcal{D}(K_0) (k_Fl_e)^{-1} (k_F\xi)^{2\nu_2}$ is the correction due the impurity scatterings (see App.C for details). In the above, $\mathcal{C}(K_0) = K_0^2/(16\pi^2(1-K_0))$ and $\mathcal{D}(K_0) = (K_0^2-1)/(8\pi(3-K_0-K_0^{-1}))$, $\nu_1=(2-2K_0)/(2-K_0^{-1})$, and $\nu_2=(3-K_0-K_0^{-1})/(2-K_0^{-1})$. It is instructive to write down the instanton action when $D$ and $g_u$ are small:
\beq
\mathcal{S}_{inst} = \frac{8\sqrt{K_r}}{\pi} \left[ \frac{L}{\xi} - \frac{L\xi}{2l_u^2} \mathcal{F}_1 - \frac{L}{2l_e} \mathcal{F}_2 \right],
\eeq
where $\mathcal{F}_1 = \mathcal{F}_1(K_0,k_F,\xi) = \mathcal{C}(K_0)(k_F\xi)^{2(\nu_1-1)}$ and $\mathcal{F}_2 = \mathcal{F}_2(K_0,k_F,\xi) = \mathcal{D}(K_0)(k_F\xi)^{\nu_2-1}$. From these expressions, we identify that the Umklapp interaction would be detrimental to the MKDs with the Luttinger parameter $K_0<1$ (to make $\mathcal{F}_1$ positive), while the disorder would be fatal with $1<K_0<(3+\sqrt{5})/2$ (to make $\mathcal{F}_2$ positive). In contrast, the disorder does harms to the Majorana fermions in a Kitaev chain, when $K<3/2$ \cite{Lobos12}. The presence of Umklapp interaction shifts the range of Luttinger parameters, in which the disorder plays negative roles. However, even when the Luttinger parameter is within the range mentioned in the above, the MKDs would be still stable, if the strengths of the Umklapp interaction and the disorder are small so that $\xi (k_F\xi)^{\nu_1-1}\ll l_u$ and $\xi (k_F\xi)^{\nu_2-1}\ll l_e$, under which conditions the contributions from the Umklapp interaction and disorder become sub-leading. Within the Bosonization framework $k_F \xi \gg 1$ would hold, so when $K_0<\sqrt{2}/2$ (to make $\nu_1>1$) and $1/2<K_0<1$ (to make $\nu_2>1$) the Umklapp interaction and disorder are much easier to reach the threshold to breaking the conditions $\xi (k_F\xi)^{\nu_1-1}\ll l_u$ and $\xi (k_F\xi)^{\nu_2-1}\ll l_e$.

The above analyses imply an interesting fact about the disorder: when $1/2<K_0<1$, a smaller disorder strength threshold is required to compete with the superconducting order, but at the same time the corresponding term in Eq.(13) changes the sign, which means that disorder does not provide negative effects on the degeneracy energy splitting; on the other hand, when $1<K_0<(3+\sqrt{5})/2$, the disorder strength threshold to compete with the superconducting order is enlarged very much because of $\nu_2<1$. Therefore, disorder does not place a serious problem for the energy splitting of MKDs. The energy splitting would be seriously affected by the Umklapp interaction, when $K_0<\sqrt{2}/2$. Within this range of Luttinger parameter, the Umklapp interaction strength threshold is reduced because $\nu_1>1$ and the sign of the corresponding term in Eq.(13) does not change. Therefore, we conclude that the Umklapp interaction and disorder have very weak influence on the degeneracy splitting, as long as $K_0>\sqrt{2}/2$.

\begin{center}
\textbf{Conclusion}
\end{center}

We have studied the effects of the interplay between interactions and disorder on $1D$ DIII topological superconductors and the MKDs supported therein. By performing the perturbative RG analysis, the stable topological regime is found to be rather wide in the parameter space formed by $\{y_\Delta,y_g,y_D,K\}$. Moreover, the topological degeneracy splitting of MKDs are also analyzed by performing the instanton analysis. It turns out that the splitting of MKDs due to the correction of the Umklapp interactions and disorders can be omitted, when $\xi (k_F\xi)^{v_1-1}\ll l_u$ and $\xi (k_F\xi)^{v_2-1}\ll l_e$. Our theory predicts that the MKDs are generally stable against both the Umklapp interaction and disorder when $K_0>\sqrt{2}/2$.

\pagebreak
\widetext

\begin{center}
\textbf{\large Appendix}
\end{center}

\setcounter{figure}{0}
\makeatletter
\renewcommand{\thefigure}{A\arabic{figure}}
\renewcommand{\bibnumfmt}[1]{[A#1]}
\renewcommand{\citenumfont}[1]{#1}

The appendix is organized as follows: the derivation of the model specified by Eq.(1) in the main text in terms of chiral Fermions around the Fermi level are provided in Sec.A. The details of the derivation of RG equations Eq.(7)-(11) in the main text are illustrated in Sec.B. The calculations of the energy splitting of MKDs in the presence of Umklapp interactions and disorder are shown in Sec.C.

\subsection{The Hamiltonian in terms of chiral Fermions}

\setcounter{equation}{0}
\renewcommand{\theequation}{A\arabic{equation}}

In $k$-space, the Hamiltonian can be written as:
\beq
H = \sum_{k} \Psi_k^\dag \left(\begin{array}{cc} \mathcal{H}_{nw}(k) & \Delta(k) \\ \Delta^\dag(k) & -\mathcal{H}_{nw}^T(-k) \end{array}\right) \Psi_k = \sum_{k} \Psi_k^\dag \mathcal{H}(k) \Psi_k,
\eeq
where $\mathcal{H}_{nw}(k)=(-2t\cos k - \mu)\sigma_0 + \lambda_R \sin k \sigma_y$, $\Delta(k)=i \Delta \cos k \sigma_y$, and the Nambu's spinor is written as:
\beq
\Psi_k=\left(\begin{array}{cccc} c_{k,\uparrow} & c_{k,\downarrow} & c_{-k,\uparrow}^\dag & c_{-k,\downarrow}^\dag \end{array}\right)^T, \nonumber
\eeq
due to the presence of Rashba SOC. It is obvious that the Hamiltonian density $\mathcal{H}(k)$ fulfills both the time-reversal and particle-hole symmetries with the two operators defined as $\mathcal{T}=\tau_0\otimes i\sigma_y \mathcal{K}$  and $\mathcal{P} = \tau_x\otimes \sigma_0 \mathcal{K}$ with $\mathcal{K}$ denoting the complex conjugate operator. In the above, $\tau$-matrices and $\sigma$-matrices act on the particle-hole and the spin space respectively.

To study the low-energy physics, we notice that in the limit $\Delta=0$ the Hamiltonian can be diagonalized by introducing:
\beq
\left(\begin{array}{c} d_{n,+} \\ d_{n,-} \end{array}\right) = \frac{1}{\sqrt{2}} \left(\begin{array}{cc} 1 & -i \\ -i & 1 \end{array}\right) \left(\begin{array}{c} c_{n,\uparrow} \\ c_{n,\downarrow} \end{array}\right).
\eeq
Then the Hamiltonian $H(\Delta=0)$ is split into two bands labeled by $\eta=\pm1$ explicitly:
\beq
H(\Delta=0) = \sum_{k,\eta} \left(-2\tilde{t}\cos(ka+\eta q_0 a)-\mu\right) d_{k,\eta}^\dag d_{k,\eta},
\eeq
where $\tilde{t}=\sqrt{t^2+\lambda_R^2}$ and $q_0a=\arctan(\lambda_R/t)$. Since we are interested in the low-energy physics, we utilize the transformation: $na \rightarrow x$ and $\sum_{n} \rightarrow \int dx/a$. Then after rewriting the Hamiltonian in the new basis and expressing the operator $d_{n,\eta}$ into the left- and right-moving excitations for each band $\eta$:
\beq
d_{n,\eta} = \sqrt{a} \left(e^{i(k-\eta q_0)x} R_{\eta}(x) + e^{i(-k-\eta q_0)x} L_{\eta}(x)\right),
\eeq
we then obtain the low-energy effective Hamiltonian in terms of the right-moving and left-moving quasiparticles:
\begin{align}
H = &-iv_F \sum_{\eta=\pm} \int dx \left[R_{\eta}^\dag(x)\partial_xR_{\eta}(x)-L_{\eta}^\dag(x)\partial_xL_{\eta}(x)\right] \nonumber \\
& + 2\Delta \cos(k_F+q_0)a \int dx \left[L_+^\dag(x) R_-^\dag(x) + h.c.\right] \nonumber \\
& + 2\Delta \cos(k_F-q_0)a \int dx \left[R_+^\dag(x) L_-^\dag(x) + h.c.\right].
\end{align}

We then consider the effect of interaction and disorder. The interactions in the system can be included completely by the analog with the 'g-ology' of spinful Luttinger liquid:
\beq
H_1 = \int dx \sum_{\tau=\pm} \left[ g_{1,\perp} R_{\tau}^\dag (x) L_\tau(x) L_{-\tau}^\dag(x) R_{-\tau}(x) + g_{1,\parallel} R_{\tau}^\dag (x) L_\tau(x) L_{\tau}^\dag(x) R_{\tau}(x) \right],
\eeq
\beq
H_2 = \int dx \sum_{\tau=\pm} \left( g_{2,\parallel} R_{\tau}^\dag(x) R_{\tau}(x) L_{\tau}^\dag(x) L_{\tau}(x) + g_{2,\perp} R_{\tau}^\dag(x) R_{\tau}(x) L_{-\tau}^\dag(x) L_{-\tau}(x) \right),
\eeq
\beq
H_3 = \int dx \sum_{\tau} \frac{g_3}{2} \left( e^{i4k_Fx}R_\tau^\dag(x) R_{-\tau}^\dag(x) L_{\tau}(x) L_{-\tau}(x) + h.c.\right),
\eeq
\beq
H_4 = \int dx \sum_{\tau} \left[ \begin{array}{l} \frac{g_{4,\parallel}}{2} \left(R_\tau^\dag(x) R_\tau(x) R_\tau^\dag(x) R_\tau(x) + L_\tau^\dag(x) L_\tau(x) L_\tau^\dag(x) L_\tau(x)\right) \\ + \frac{g_{4,\perp}}{2} \left(R_\tau^\dag(x) R_{\tau}(x) R_{-\tau}^\dag(x) R_{-\tau}(x)  + L_\tau^\dag(x) L_\tau(x) L_{-\tau}^\dag(x) L_{-\tau}(x)\right) \end{array} \right].
\eeq
The first term $H_1$ is known as the backward scattering term with the coupling strength $g_{1,\perp}$ and $g_{1,\perp}$. As its name states, this term scatters a left-moving (right-moving) electron to the right-moving (left-moving) so the direction of the electron propagating changes. The second term $H_2$ is called as the dispersive scattering with coupling strength $g_{2,\perp}$ and $g_{2,\parallel}$. It scatters the particles within its own channel. The third term $H_3$ is known as the Umklapp scattering. Two particles changes its channel. Obviously, this would acquire a phase $e^{i4k_Fx}$ (details see $161484$ page 11), so this term would be not important unless the system is close to half-filling. For our case, the chemical potential of interest is just around half-filling (the center of the band with a window of width $\lambda_{so}$), so we expect that this term would be important. The fourth term $H_4$ is known as the forward scattering term. The particles scatters within the same channel.

In the low-energy limit, we just focus on what happens around Fermi points. In general, the impurity Hamiltonian can be written as:
\beq
H_{dis} = \int dx V(x) \rho(x),
\eeq
where $\rho(x)$ is the Fermi density at location $x$, and $V(x)$ is the scattering potential and fulfills $\langle V(x) V(y)\rangle=D\delta(x-y)$. Since we are interested in the physics around Fermi points, we can write Eq.(N55) as:
\beq
H_{dis} = \frac{1}{N} \sum_{\tau,q\sim 0} V_q \sum_{k} d_{k+q,\tau}^\dag d_{k,\tau} + \frac{1}{N} \sum_{\tau,q\sim\pm 2k_F} V_q \sum_{k} d_{k+q,\tau}^\dag d_{k,\tau},
\eeq
where we have assumed that the impurities are usual scalar potentials and can not induce scattering between different channels. Then using the right-moving and left moving Fermions, we have:
\beq
H_{dis} = \sum_{\tau} \int dx \left[\eta(x) \left(R_{\tau}^{\dag}(x) R_{\tau}(x) + L_{\tau}^{\dag}(x)L_{\tau}(x)\right) + \left(\xi(x) e^{i2k_Fx} L_{\tau}^{\dag}(x) R_{\tau}(x) + \xi^*(x) e^{-i2k_Fx} R_{\tau}^{\dag}(x) L_{\tau}(x)\right)\right].
\eeq
In these terms, the term $\sim g_{1,\perp}$ is usually known to be irrelevant, we thus ignore it in the study. The term $\sim g_{1,\parallel}$ can be absorbed into $g_{2,\parallel}$, which just lead to the redefinition of $g_{2,\parallel}$. In the disorder part of the Hamiltonian, the forward scattering can be integrated out, so we also drop this term.

With the Hamiltonian in terms of chiral fermions given by Eq.(A5)-(A9) and Eq.(A12), we can follow the standard Bosonization procedure \cite{Giamarchi04} to obtain Eq.(4) in the main text.

\subsection{The derivation of RG equations}

\setcounter{equation}{0}
\renewcommand{\theequation}{B\arabic{equation}}

We just begin with the low-energy effective action given by Eq.(4) in the main text:
\begin{align}
S &= \sum_{\mu}\int dx d\tau \left[-\left(\mathcal{L}_1^\mu + \mathcal{L}_2^\mu\right) + \frac{\tilde{\Delta}}{\pi \alpha} \sin2\vartheta_1^\mu - \frac{\tilde{\Delta}}{\pi \alpha} \sin2\vartheta_2^\mu + \frac{g_u}{2\pi^2 \alpha^2} \cos 2(\varphi_1^\mu(x) + \varphi_2^\mu(x))\right] \nonumber \\
&-\frac{D}{\pi^2 \alpha^2} \sum_{\mu,\nu} \int dx d\tau d\tau' \left[\begin{array}{c} \cos\left(\vartheta_1^\mu(x,\tau)+\vartheta_2^\mu(x,\tau)\right) \cos\left(\vartheta_1^\nu(x,\tau')+\vartheta_2^\nu(x,\tau')\right) \\ \cos\left(\varphi_1^\mu(x,t)+\varphi_2^\mu(x,t)-\varphi_1^\nu(x,\tau')-\varphi_2^\nu(x,\tau')\right) \end{array}\right],
\end{align}
where we take $\tilde{\Delta}=\Delta\cos(k_F+q_0)a$ and $\tilde{\Delta}=-\Delta\cos(k_F-q_0)a$. The obvious symmetry is the exchange of fields labeled by $1$ with those labeled by $2$. Therefore, we just need to calculate the following correlation function:
\beq
R_{\vartheta_1}(x_2-x_1,\tau_2-\tau_1) = \left\langle T_{\tau} e^{i\vartheta_{1}(x_1,\tau_1)} e^{-i\vartheta_{1}(x_2,\tau_2)} \right\rangle,
\eeq
to obtain the flow equations. We then have to expand the action up to the order of $\tilde{\Delta}^2$, $g_u^2$, and $D$ to analyze the effect from these terms on the ground states.

For the term with $\tilde{\Delta}^2$, we can write it as:
\beq
\mathcal{T}_1 = \frac{-1}{2}\left(\frac{\tilde{\Delta}}{2\pi \alpha}\right)^2 \int dx_3 d\tau_3 dx_4 d\tau_4 \sum_{\epsilon_3,\epsilon_4=\pm1} \left\langle \begin{array}{l} T_{t} e^{i\vartheta_1(x_1,\tau_1)} e^{-i\vartheta_1(x_2,\tau_2)} \epsilon_3 e^{i\epsilon_3 2\vartheta_1(x_3,\tau_3)} \epsilon_4 e^{i\epsilon_4 2\vartheta_1(x_4,\tau_4)} \end{array} \right\rangle.
\eeq
The factor $-1$ in red is due to the multiplication of $i$ in the $\sin$ function. The straightforward calculation yields:
\beq
\mathcal{T}_1(\vec{r}_1-\vec{r}_2) = \frac{\tilde{\Delta}^2 F_1(\vec{r}_1-\vec{r}_2)}{2 u^2 K^2 \alpha^2} e^{-\frac{1}{2K}F_1(\vec{r}_1-\vec{r}_2)}  \int dr r^3 e^{-\frac{2F_1(\vec{r})}{K}} = \frac{\tilde{\Delta}^2 F_1(\vec{r}_1-\vec{r}_2)}{2 u^2 K^2 \alpha^2} e^{-\frac{1}{2K}F_1(\vec{r}_1-\vec{r}_2)}  \int dr r^3 \left(\frac{r}{\alpha}\right)^{-\frac{2}{K}}.
\eeq

For the term $\sim g_u^2$, we denote these as:
\beq
\mathcal{T}_2 = \frac{1}{2} \left(\frac{g_u}{4\pi^2\alpha^2}\right)^2 \int dx_3 d\tau_3 dx_4 d\tau_4 \sum_{\epsilon_3,\epsilon_4=\pm1} \left\langle \begin{array}{l} T_{t} e^{i \vartheta_1(x_1,\tau_1)} e^{-i \vartheta_1(x_2,\tau_2)} e^{i\epsilon_3 2\varphi_1(x_3,\tau_3)} \\ e^{i\epsilon_4 2\varphi_1(x_4,\tau_4)} e^{i\epsilon_3 2\varphi_2(x_3,\tau_3)} e^{i\epsilon_4 2\varphi_2(x_4,\tau_4)} \end{array} \right\rangle.
\eeq
After some calculations, we find:
\beq
\mathcal{T}_2 (\vec{r}_1-\vec{r}_2) = \frac{-g_u^2 F_1(\vec{r}_1-\vec{r}_2)}{8\pi^2 u^2} e^{-\frac{1}{2K} F_1(\vec{r}_1-\vec{r}_2)} \int \frac{dr}{\alpha} \left(\frac{r}{\alpha}\right)^{3-4K}.
\eeq

For the term $\sim D$, we denote it as:
\beq
\mathcal{T}_3 = \frac{D}{8\pi^2 \alpha^2} \int dx_3 d\tau_3 dx_4 d\tau_4 \delta(x_3-x_4) \sum_{\epsilon_3,\epsilon_4,\epsilon_5=\pm1} \left\langle \begin{array}{l} T_{t} e^{i \vartheta_1(x_1,\tau_1)} e^{-i \vartheta_1(x_2,\tau_2)} e^{i\epsilon_3 \vartheta_1(x_3,\tau_3)} e^{i\epsilon_3 \vartheta_2(x_3,\tau_3)} \\ e^{i\epsilon_4 \vartheta_1(x_4,\tau_4)} e^{i\epsilon_4 \vartheta_2(x_4,\tau_4)} \\ e^{i\epsilon_5 \varphi_1(x_3,\tau_3)} e^{i\epsilon_5 \varphi_2(x_3,\tau_3)} e^{-i\epsilon_5 \varphi_1(x_4,\tau_4)} e^{-i\epsilon_5 \varphi_2(x_4,\tau_4)} \end{array} \right\rangle.
\eeq
After evaluating the averages, we would find that:
\beq
\mathcal{T}_3(\vec{r}_1-\vec{r}_2) = \frac{D \alpha}{16\pi u^2} e^{-\frac{1}{2K} F_1(\vec{r}_1-\vec{r}_2)} \left( \frac{1}{K^2}-1 \right) \left[F_1(\vec{r}_1-\vec{r}_2) - \frac{\cos \theta_{\widehat{\vec{r}_1 \vec{r}_2}}}{2} \right]  \int \frac{dy}{\alpha} \left(\frac{y}{\alpha}\right)^{2-(K+K^{-1})}
\eeq
where $\theta_{\widehat{\vec{r}_1 \vec{r}_2}}$ denotes the angle between two vectors $\vec{r}_1$ and $\vec{r}_2$.

With the above results we can obtain the renormalization flow equations. In the processes of RG, we find that a term $\propto \cos \theta_{\widehat{\vec{r}_1 \vec{r}_2}}$ is generated. Therefore, we should modify the definition of $F_1$:
\beq
F_1(\vec{r}) = \frac{1}{2}\ln \frac{x^2 + u^2\tau^2}{\alpha^2} + \gamma \cos \theta_{\widehat{\vec{r}}} = I(\vec{r}) + \gamma \cos \theta_{\widehat{\vec{r}}}.
\eeq
Therefore, we would have the expression of $R_{\vartheta_1}$ as:
\begin{align}
R_{\vartheta_1} \left( \vec{r}_1-\vec{r}_2 \right) = e^{-\frac{1}{2K} F_1(\vec{r}_1-\vec{r}_2)} \left[ \begin{array}{l} 1 + \frac{\tilde{\Delta}^2 \alpha^2 I(\vec{r}_1-\vec{r}_2)}{2 K^2 u^2} \int \frac{dr}{\alpha} \left(\frac{r}{\alpha}\right)^{3-\frac{2}{K}} - \frac{g_3^2 I(\vec{r}_1-\vec{r}_2)}{8\pi^2 u^2} \int \frac{dr}{\alpha} \left(\frac{r}{\alpha}\right)^{3-4K} \\
+ \frac{D \alpha I(\vec{r}_1-\vec{r}_2)}{16 \pi u^2} \left( \frac{1}{K^2}-1 \right) \int \frac{dy}{\alpha} \left(\frac{y}{\alpha}\right)^{2-(K+K^{-1})} \\
- \frac{D \alpha \cos \theta_{\widehat{\vec{r}_1 \vec{r}_2}}}{32 \pi u^2} \left( \frac{1}{K^2}-1 \right) \int \frac{dy}{\alpha} \left(\frac{y}{\alpha}\right)^{2-(K+K^{-1})}
\end{array} \right].
\end{align}
Then we would have:
\beq
\begin{cases}
\frac{1}{K_{eff}} = \frac{1}{K} - \frac{y_\Delta^2 }{ K^2} \int \frac{dr}{\alpha} \left(\frac{r}{\alpha}\right)^{3-\frac{2}{K}} + \frac{y_g^2}{4} \int \frac{dr}{\alpha} \left(\frac{r}{\alpha}\right)^{3-4K} - \frac{y_D}{8} \left( \frac{1}{K^2}-1 \right) \int \frac{dy}{\alpha} \left(\frac{y}{\alpha}\right)^{2-(K+K^{-1})}, \\
\gamma_{eff} = \gamma + \frac{y_D}{16} \left( \frac{1}{K}-K  \right) \int \frac{dy}{\alpha} \left(\frac{y}{\alpha}\right)^{2-(K+K^{-1})},
\end{cases}
\eeq
where we defined $y_\Delta = \tilde{\Delta}\alpha/u$, $y_g = g_u/\pi u$, and $y_D = D\alpha/\pi u^2$. Then we have the following RG flow equations:
\begin{align}
\frac{dy_\Delta}{d\ell} &= \left(2-K^{-1}\right)y_\Delta, \\
\frac{dy_g}{d\ell} &= \left(2-2K\right)y_g, \\
\frac{dy_D}{d\ell} &= \left(3-(K^{-1}+K)\right)y_D, \\
\frac{dK^{-1}}{d\ell} &= -\frac{y_\Delta^2}{K^2} + \frac{y_g^2}{4} - \frac{y_D}{8}\left(\frac{1}{K^2}-1\right), \\
\frac{d\gamma}{d\ell} &= \frac{y_D}{16}\left(\frac{1}{K}-K\right).
\end{align}
Moreover, we notice that:
\beq
\frac{1}{2u} \frac{du}{d\ell} = - \frac{d\gamma}{d\ell},
\eeq
Therefore, the last equation gives the renormalization of Fermi velocity:
\beq
\frac{du}{d\ell} = \frac{y_D}{8}\left(K-\frac{1}{K}\right) u.
\eeq
Substitute Eq.(B16) by Eq.(B18), we obtain the RG flow equations given in the main text.

\subsection{The calculation of energy splitting of MKDs}

\setcounter{equation}{0}
\renewcommand{\theequation}{C\arabic{equation}}

To begin with the discussion, we may rewrite the action into the following form:
\beq
\begin{cases}
S_1^{(0)} = \int dx d\tau \left[\frac{\partial_x \varphi_1}{i\pi} \partial_\tau \vartheta_1 + \frac{u}{2\pi K} (\partial_x \varphi_1)^2 + \frac{u K}{2\pi} (\partial_x \vartheta_1) + \frac{\tilde{\Delta}}{\pi \alpha} \sin 2\vartheta_1\right], \\
S_2^{(0)} = \int dx d\tau \left[\frac{\partial_x \varphi_2}{i\pi} \partial_\tau \vartheta_2 + \frac{u}{2\pi K} (\partial_x \varphi_2)^2 + \frac{u K}{2\pi} (\partial_x \vartheta_2) - \frac{\tilde{\Delta}}{\pi \alpha} \sin 2\vartheta_2\right], \\
S_{um} = \frac{g_u}{2\pi^2 \alpha^2} \int dx d\tau \cos\left[2(\varphi_1(x)+\varphi_2(x))\right], \\
S_{dis} = -\frac{D}{\pi^2 \alpha^2} \int dx d\tau d\tau' \left[ \begin{array}{c} \cos\left(\vartheta_1(x,\tau)+\vartheta_2(x,\tau)\right) \cos\left(\vartheta_1(x,\tau')+\vartheta_2(x,\tau')\right) \\ \cos\left(\varphi_1(x,\tau)+\varphi_2(x,\tau) - \varphi_1(x,\tau')-\varphi_2(x,\tau') \right) \end{array} \right].
\end{cases}
\eeq
We are interested in the topological regime where the superconducting order is strong. Then we can assume that the term proportional to $\tilde{\Delta}$ should be pinned to the minimum. For the part with the field $\vartheta_1$ we will have $\vartheta_1=\{-\pi/4,3\pi/4\}$, while for the part with the field $\vartheta_2$ we will have $\vartheta_2=\{-3\pi/4,\pi/4\}$. Then we need to consider how the two sets of minima connects.

\subsubsection{the free parts $S_1^{(0)}$ and $S_2^{(0)}$}

We first consider the case in the absence of disorder and Umklapp terms. For this purpose, we can integrate out the $\varphi$ field and assume that the $\vartheta$ locates at one of the minima uniformly in space. Since the minima of $\vartheta_1$ and $\vartheta_2$ is related by the time-reversal, we may focus on one of them. Then the action is written as:
\beq
S^{(0)} = \int dx d\tau \left[\frac{K}{2\pi u} ( \partial_\tau \vartheta)^2 + \frac{u K}{2\pi} ( \partial_x \vartheta)^2 + \frac{\tilde{\Delta}}{\pi \alpha} \sin2\vartheta\right].
\eeq
The classical motion is determined by the variation of action with respect to $\varphi$, which gives:
\beq
\frac{K}{\pi u} \partial_\tau^2 \vartheta + \frac{uK}{\pi} \partial_x^2 \vartheta = -\frac{2\tilde{\Delta}}{\pi \alpha} \cos 2\vartheta. \Rightarrow \partial_\tau^2 \vartheta + \frac{2 \tilde{\Delta} u}{K \xi} \cos2\vartheta=0.
\eeq
In the last step, we set $\alpha\rightarrow \xi$, when the superconducting correlation flows to the strong coupling. The classical instanton is given by:
\beq
\begin{cases}
\vartheta_1^{inst} = \frac{\pi}{4} + 2\arctan\left[\tanh(\tau/\tau_0)\right], \\
\vartheta_2^{inst} = -\frac{\pi}{4} + 2\arctan\left[\tanh(\tau/\tau_0)\right],
\end{cases}
\eeq
where  $\tau_0 = \sqrt{K}\xi/u$. Now we just need to find the action of the instanton:
\beq
S^{(0)} = \frac{uKL}{2\pi} \int d\tau \left[\frac{1}{u^2} (\partial_\tau \vartheta)^2 + \frac{2\tilde{\Delta}}{\xi Ku}\sin2\vartheta\right]=\frac{KL}{\pi} \int dz \left[\frac{1}{2}(\partial_z \vartheta)^2 + \frac{1}{\xi^2 K}\sin2\vartheta\right].
\eeq
The equation of motion is given by:
\beq
-\partial_z^2 \vartheta + V'(\vartheta)=0 \Rightarrow  \frac{-(\partial_z\vartheta)^2}{2} + V(\vartheta)=E.
\eeq
where $E$ is a constant depending on the initial condition. Then, we would have:
\beq
S^{(0)} = \frac{KL}{\pi} \int dz (\partial_z \vartheta)^2 = \frac{KL}{\pi} \int d\vartheta (\partial_z \vartheta) = \frac{KL}{\pi} \int_{-\pi/4}^{3\pi/4} d\vartheta \sqrt{2(V(\vartheta)-E)} = \frac{4KL}{\pi} \sqrt{\frac{1}{\xi^2 K}} = \frac{4\sqrt{K}}{\pi}\frac{L}{\xi},
\eeq
where in the first equality we dropped the trivial constant $E=\frac{1}{\xi^2 K}$. The analysis can be equally applied to $S_1^{(0)}$ and $S_2^{(0)}$, so we would have:
\beq
S_1^{(0)} = S_2^{(0)} = \frac{4\sqrt{K}}{\pi}\frac{L}{\xi}.
\eeq

\subsubsection{The correction from the term $\propto D$}

Before we do the calculations, we first rewrite the action for the disorder part:
\begin{align}
S_{dis} = &-\frac{D}{2\pi^2 \alpha^2} \int dx d\tau d\tau' \left[\begin{array}{c} \cos\left( \vartheta_1(x,\tau)+\vartheta_2(x,\tau) + \vartheta_1(x,\tau')+\vartheta_2(x,\tau') \right) \\ \cos\left(\varphi_1(x,\tau)+\varphi_2(x,\tau) - \varphi_1(x,\tau')-\varphi_2(x,\tau') \right)\end{array}\right] \nonumber \\
& -\frac{D}{2\pi^2 \alpha^2} \int dx d\tau d\tau' \left[\begin{array}{c} \cos\left( \vartheta_1(x,\tau)+\vartheta_2(x,\tau) - \vartheta_1(x,\tau')-\vartheta_2(x,\tau') \right) \\ \cos\left(\varphi_1(x,\tau)+\varphi_2(x,\tau) - \varphi_1(x,\tau')-\varphi_2(x,\tau') \right)\end{array}\right].
\end{align}
In the topological regime, $\vartheta_i$ are pinned and the time scale of a transition from one minimum to the other is $\tau_0\sim\xi$, which is very small. Analog to the argument of dilute instanton gas, inserting all the possible pinned values into the equation above, we find that the terms containing $\vartheta_i$ are vanishing. Therefore, the action is simplified as:
\beq
S_{dis} \approx -\frac{D}{\pi^2 \alpha^2} \int dx d\tau d\tau' \cos\left(\varphi_1(x,\tau)+\varphi_2(x,\tau) - \varphi_1(x,\tau')-\varphi_2(x,\tau')\right).
\eeq
Then the calculation below is parallel to that by Lobos \textit{et al} \cite{}. To evaluate the action above, we notice the following identity:
\begin{align}
&\cos\left(\varphi_1(x,\tau)+\varphi_2(x,\tau) - \varphi_1(x,\tau')-\varphi_2(x,\tau')\right) \nonumber \\
= &: \cos\left(\varphi_1(x,\tau)+\varphi_2(x,\tau) - \varphi_1(x,\tau')-\varphi_2(x,\tau')\right) : \exp\left[-\frac{1}{2}\left\langle \left[\varphi_1(x,\tau)+\varphi_2(x,\tau) - \varphi_1(x,\tau')-\varphi_2(x,\tau')\right]^2 \right\rangle_0\right].
\end{align}
Now we introduce the center of mass and relative coordinate: $\tau_c=(\tau+\tau')/2$ and $\tau_r = \tau-\tau'$. Therefore,
\beq
\varphi_1(x,\tau)+\varphi_2(x,\tau) - \varphi_1(x,\tau')-\varphi_2(x,\tau') = \partial_{\tau_c} \varphi_1(x,\tau_c) \tau_r + \partial_{\tau_c} \varphi_2(x,\tau_c) \tau_r.
\eeq
As a result, Eq.(C11) reduces to:
\begin{align}
&\cos\left(\varphi_1(x,\tau)+\varphi_2(x,\tau) - \varphi_1(x,\tau')-\varphi_2(x,\tau')\right) \nonumber \\
\approx & \left(1 - \frac{1}{2} (\partial_{\tau_c} \varphi_1(x,\tau_c) + \partial_{\tau_c} \varphi_2(x,\tau_c) )^2 \tau_r^2\right) \exp\left[-\frac{1}{2}\left[\left\langle (\varphi_1(x,\tau) - \varphi_1(x,\tau'))^2 \right\rangle_0+\left\langle(\varphi_2(x,\tau) -\varphi_2(x,\tau'))^2\right\rangle_0\right] \right].
\end{align}
$\partial_{\tau_c} \varphi_1(x,\tau_c)$ can be evaluated from the equation of motion, and the results are:
\beq
\begin{cases}
\dot{\varphi}_1(x,\tau_c) = -iv(\ell^*) K(\ell^*) \nabla\vartheta_1(x,\tau_c) + \frac{i2\pi\Delta(\ell^*)}{\xi} \int dx' \textrm{sgn}(x'-x) \sin 2\vartheta_1(x',\tau_c), \\
\dot{\varphi}_2(x,\tau_c) = -iv(\ell^*) K(\ell^*) \nabla\vartheta_2(x,\tau_c) - \frac{i2\pi\Delta(\ell^*)}{\xi} \int dx' \textrm{sgn}(x'-x) \sin 2\vartheta_2(x',\tau_c).
\end{cases}
\eeq
Inserting Eq.(C14) into Eq.(C13) and integrating over $\tau_c$ we find:
\beq
S_{dis} = -\frac{32D(\ell^*) \tau_0(\ell^*)}{3\xi^2} \left( \frac{\Delta(\ell^*)}{\xi} \right)^2 \int_{-L/2}^{L/2} dx x^2 \int d\tau_r \tau_r^2 e^{-\frac{1}{2}\left[\left\langle (\varphi_1(x,\tau) - \varphi_1(x,\tau'))^2 \right\rangle_0+\left\langle(\varphi_2(x,\tau) -\varphi_2(x,\tau'))^2\right\rangle_0\right]}.
\eeq
In the calculation above we have dropped the trivial constant from the first term in the round bracket of Eq.(C13). The correlation of $\varphi_i$ in the exponential evaluated under the free action $S_i^{(0)}$ is given by \cite{Lobos12}:
\beq
\left\langle [\varphi_i(\tau)-\varphi_i(0)]^2 \right\rangle = 2\pi \left[ \frac{x^2}{L^2} + \frac{L}{\pi^2} \right] \sqrt{\frac{\Delta(\ell^*) K(\ell^*)}{v(\ell^*)\xi}} \left(1-\exp(-2|\tau|/\tau_0)\right).
\eeq
Inserting this into Eq.(C15) and the integration finally gives $S_{dis} \sim \frac{\xi}{\ell_e}$ \cite{Lobos12}. This term can be ignored, as we consider the thermodynamic limit with $L/\xi\rightarrow\infty$.

\subsubsection{The correction from the term $\propto g_u$}

The energy splitting due to this term is given by:
\beq
\delta E_{g} \propto \exp\left[-\frac{g_u}{2\pi^2 \alpha^2} \int dx d\tau \cos\left[2(\varphi_1(x)+\varphi_2(x))\right]\right].
\eeq
We can expand the exponential in the powers of $g_u$ due to the fact $g_u\ll1$. In the strong coupling of $y_{\Delta}$, the fields $\vartheta_1$ and $\vartheta_2$ are pinned. The conjugated fields $\varphi_1$ and $\varphi_2$ are highly fluctuating. Moreover, $\varphi_1$ and $\varphi_2$ are independent, and thus the term $\cos(\varphi_1(x,\tau)+\varphi_2(x,\tau))$ is also highly fluctuating in space and time. This leads to the vanishing of the Umklapp action $S_{um}$ in the leading order of $g_u$.

\subsubsection{The indirect contribution from the Luttinger parameter}

Finally we calculate the renormalization of $K(\ell^*)$. By using the lowest order approximation for $K(\ell)=K_0$, from the RG equations for $y_\Delta$, $y_g$, and $y_D$ we would have: $y_\Delta(\ell) = y_{\Delta,0} e^{(2-K_0^{-1})\ell}$, $y_{g}(\ell)=y_{g,0} e^{(2-2K_0)\ell}$, and $y_D = y_{D,0}e^{[3-(K_0^{-1}+K_0)]\ell}$. In the regime we are interested in, $y_\Delta$ first goes to the strong coupling. From this fact, we find $\ell^* = \ln(y_{\Delta,0})/(K_0^{-1}-2)$. From this and the RG equation for $K$, we find $K(\ell)$ in the lowest order approximation:
\beq
K(\ell^*) = K_0 + \int_0^{\ell^*} d\ell' \frac{dK(\ell')}{d\ell'}.
\eeq
It is very illustrative to write $\tilde{\Delta}=u/\xi$, $D=u^2/l_{e}$ and $g_u=u\alpha/l_u$. In the above, $\xi$ is the superconducting coherence length, $l_e$ is the scattering length, and $l_u$ is the length scale associated with Umklapp scattering. Then we would have:
\beq
y_\Delta = \frac{\alpha}{\xi},~ y_g = \frac{\alpha}{\pi l_u},~y_D = \frac{\alpha}{\pi l_e}.
\eeq
Therefore, we finally have:
\beq
K(\ell^*) = K_r - \delta K_{g} - \delta K_{D},
\eeq
where
\beq
K_r = K_0 + \frac{K_0}{4K_0-2},
\eeq
\beq
\delta K_{g}=\frac{K_0^2}{16(1-K_0)} \left( \frac{y_{g,0}}{y_{\Delta,0}^{\nu_1}} \right)^2 = \frac{K_0^2}{16\pi^2(1-K_0)} \left(\frac{\alpha}{l_u}\right)^2 \left(\frac{\xi}{\alpha}\right)^{2\nu_1},
\eeq
and
\beq
\delta K_{D}=\frac{K_0^2-1}{24-8(K_0^{-1}+K_0)} \frac{y_{D,0}}{y_{\Delta,0}^{\nu_2}} = \frac{K_0^2-1}{24\pi-8\pi(K_0^{-1}+K_0)} \left(\frac{\alpha}{l_e}\right) \left(\frac{\xi}{\alpha}\right)^{\nu_2}.
\eeq
In the above, $K_r$ is the renormalized Luttinger parameter, $\delta K_{g}$ is the correction from the Umklapp term, and $\delta K_{D}$ is the correction from the disorders.

\end{document}